\documentclass[aps,prb,twocolumn,10pt,floatfix,superscriptaddress]{revtex4-1}
\usepackage{ucs}
\usepackage[utf8x]{inputenc}  
\usepackage[T1]{fontenc}    
\usepackage[english]{babel}
\usepackage{graphicx}
\usepackage{amsmath}

 \usepackage{color}

\begin{document}
\title{Direct Measurement of Kramers' Turnover with a Levitated Nanoparticle}
\author{L.~Rondin}
\altaffiliation{Present address: Laboratoire Aim\'{e} Cotton, CNRS, Universit\'{e} Paris-Sud, ENS Cachan, Universit\'{e} Paris-Saclay, 91405 Orsay Cedex, France}
\affiliation{ETH Z\"urich, Photonics Laboratory, 8093 Z\"urich, Switzerland} 
\author{J.~Gieseler}
\altaffiliation{Present address:Physics Department, Harvard University, Cambridge, Massachusetts 02318, USA.}
\affiliation{ETH Z\"urich, Photonics Laboratory, 8093 Z\"urich, Switzerland} 
\author{F.~Ricci}
\affiliation{ICFO-Institut de Ciencies Fotoniques, The Barcelona Institute of Science and Technology, 08860 Castelldefels (Barcelona), Spain}
\author{R.~Quidant}
\affiliation{ICFO-Institut de Ciencies Fotoniques, The Barcelona Institute of Science and Technology, 08860 Castelldefels (Barcelona), Spain}
\affiliation{ICREA–Instituci\'o Catalana de Recerca i Estudis Avan\c cats, 08010 Barcelona, Spain}
\author{C.~Dellago}
\affiliation{University of Vienna, Faculty of Physics, Boltzmanngasse 5, 1090 Wien, Austria}
\author{L.~Novotny}
\email{http://www.photonics.ethz.ch}
\affiliation{ETH Z\"urich, Photonics Laboratory, 8093 Z\"urich, Switzerland}

\maketitle

\textbf{
 Understanding the thermally activated escape from a metastable state is at the heart of important phenomena such as the folding dynamics of proteins~\cite{best06,chung15}, the kinetics of chemical reactions~\cite{garcia-mueller08} or the stability of mechanical systems~\cite{badzey05}.  In 1940 Kramers calculated escape rates both in the high damping and the low damping regimes and suggested that the rate must have a maximum for intermediate damping~\cite{kramers40}. This phenomenon, today known as the Kramers turnover, has triggered important theoretical and numerical studies~\cite{haenggi90}. However, to date there is no direct and quantitative experimental verification of this turnover. Using a nanoparticle trapped in a bi-stable optical potential we experimentally measure the nanoparticle's transition rates for variable damping and directly resolve the Kramers turnover. Our measurements are in agreement with an analytical model that is free of adjustable parameters. The levitated nanoparticle presented here is a versatile experimental platform for studying and simulating a wide range of stochastic processes and testing theoretical models and predictions.}

\noindent
The Kramers turnover is relevant in many  fields of study, including the folding dynamics of proteins~\cite{best06,chung15}, the kinetics of chemical reactions~\cite{garcia-mueller08}, or current switching in Josephson tunnel junctions~\cite{silvestrini88}.
However, despite its long history and extensive theoretical work, experimental attempts to observe the Kramers turnover remain sparse. For example, experiments involving Josephson tunnel junctions verified Kramers' theoretical predictions in the underdamped regime~\cite{turlot89} and highlighted the existence of different damping regimes, but systematic errors prevented precise measurements of the transition rates in the turnover region~\cite{silvestrini88}. Kramers' theory has also been investigated using molecular isomerisation, where  the damping was controlled through the solvent density~\cite{schroeder95}. However, such experiments could only be performed in a limited damping range, with a restricted set of parameters, and no absolute knowledge about the potential energy surface, which  prevented a quantitative analysis and an unambiguous experimental verification of Kramers' turnover. 
In contrast, experiments using trapped particles in water provided quantitative results~\cite{mccann99}, but only for the overdamped case, and without resolving the Kramers' turnover.
Inspired by recent trapping experiments in vacuum~\cite{li10,gieseler12,gieseler14,millen14} we extend previous measurements into the underdamped regime for a particle in a double well trap. Since the damping coefficient $\Gamma$ of a nanoparticle is proportional to the buffer gas pressure $P_\mathrm{gas}$~\cite{gieseler12}, we are able to tune the system's damping by several orders of magnitude  by adjusting the gas pressure. This control, together with an accurate knowledge of the trapping potential, allows us to determine the rates for well-to-well transitions in both the underdamped and overdamped regimes for the same particle, thereby experimentally revealing Kramers' turnover for the first time. \\[-2.5ex]

\begin{figure}
    \begin{center}
        \includegraphics[width=.5\textwidth]{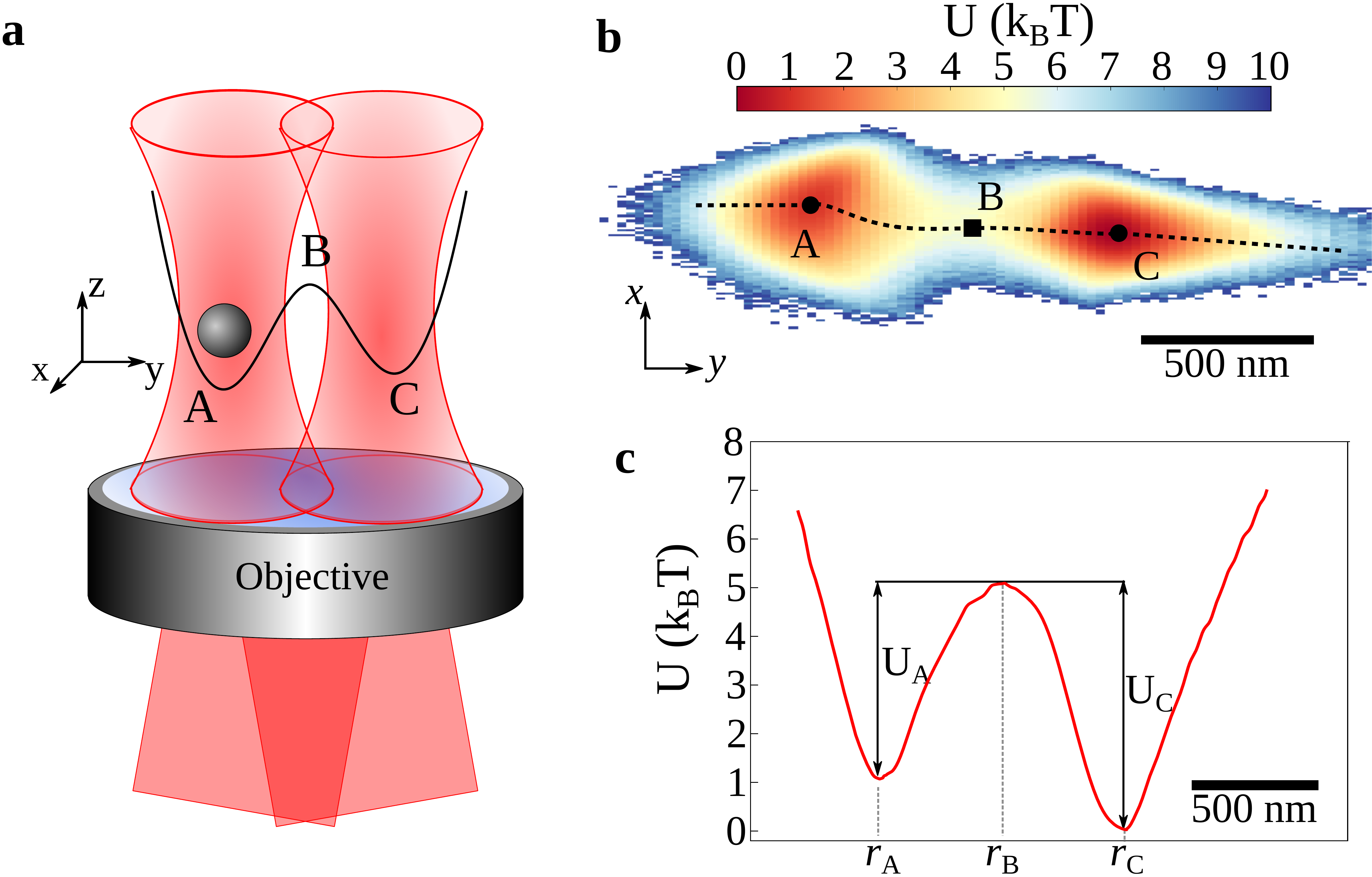}
    \end{center}
    \caption{Double-trap setup. \textbf{a}, Two focused infrared beams generate an optical potential with two wells (A and C), linked by a saddle point $B$. The experiment is operated inside a vacuum chamber. \textbf{b}, 2D-cross-section of the reconstructed optical potential in the transverse $(x,y)$ plane. The double-well potential has two stable stationary points $A$ and $C$, and an unstable stationary point $B$. The dotted line represents the minimum energy path between points $A$ and $C$. \textbf{c} Trapping potential evaluated along the dotted line in \textbf{b}. The potential energy barriers $U_A$ and $U_C$ are resolved clearly.
    \label{fig:Pot}}
\end{figure}
%

\begin{figure*}
    \begin{center}
        \includegraphics[width=.85\textwidth]{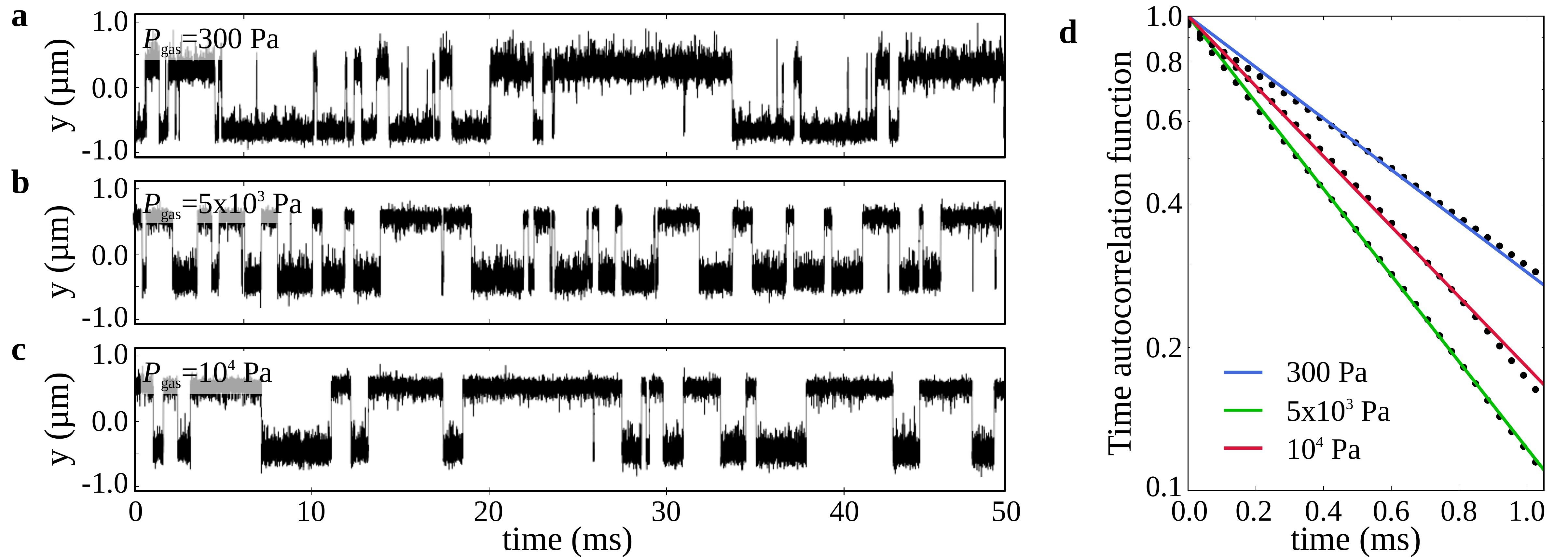}
    \end{center}
    \caption{\textbf{a-c}, Time traces along the double trap $y$ axis taken at different pressures. \textbf{d}, Time autocorrelation functions of the population operators,  $\langle \delta h_A(0) \delta h_A(t) \rangle $, obtained from the time traces. These time autocorrelations are fitted (solid color lines) by an exponential decay, allowing us to extract the jumping rate constant $R$, for each pressure. 
    \label{fig:Rates}}
\end{figure*}
The experimental configuration is illustrated in figure~\ref{fig:Pot}\textbf{a}. 
A 68~nm radius silica particle is trapped in an optical double-well potential inside a vacuum chamber of variable pressure. The potential is formed by a pair of strongly focussed infrared laser beams. 
The particle's three-dimensional trajectory is measured interferometrically with balanced split detection~\cite{gieseler12} using an additional weakly focused $\lambda=532$~nm laser beam (see Supplementary Section~A). This approach allows us to resolve the fast dynamics of the particle  (typically on the order of 100~kHz) and measure its time-dependent position $\mathbf r(t)$ with high accuracy.
Particle position time traces are recorded at $2\,$MSamples/s, for 120~s, which is longer than the thermalisation time $1/\Gamma$ for all dampings considered in this work. Therefore, the distribution of particle positions $\rho(\mathbf r)$, which we obtain by histogramming the timetraces, is thermalised. We can, thus, retrieve the optical potential $U(\mathbf r)$ from the relation $\rho(\mathbf r) = Z^{-1}\exp[{-U(\mathbf r)}/{k_B T}]$, where $Z$ is a normalisation constant,  $k_B$ the Boltzmann constant, and  $T$ is the bath temperature (see Methods).

One challenge of the present experiment compared to previous experiments involving optically levitated nanoparticles is the large volume visited by the particle (typically 1~$\mu$m$^3$), which makes the measurement scheme highly nonlinear. To obtain an accurate description of the potential we first characterise the detection scheme and account for nonlinearities present for large deviations of particle position from the origin (see Supplementary Section~B). The reconstructed potential corresponds to the actual potential experienced by the particle, accounting for any misalignments as well as possible perturbations arising from the measurement beam. 
As an example, a 2D-cross-section of the measured potential is shown in figure~\ref{fig:Pot}\textbf{b}, highlighting the presence of two stable wells $A$ and $C$ (black dots in the figure) and of one saddle point $B$ (black square). 
Important physical parameters for the description of the particle dynamics can be extracted from the measured optical potential. 
Near the stable and saddle points the potential is locally harmonic. Thus, a local harmonic fit around these points provides the characteristic frequencies $\omega_i^A$,  $\omega_i^B$ and  $\omega_i^C$, where the index $i$ indicates the local principal axis, ($i\in\{x',y',z'\}$)~\cite{mccann99}. 

Among the three characteristic frequencies at the saddle point $\omega_{y'}^B$ is pure imaginary and is denoted $\omega_S^B$ in the following given its special interest.
This curvature is associated with an unstable normal mode, and determines the transition from the underdamped to the overdamped regime.
Roughly, the turnover is expected to occur for a damping at which the energy dissipated when moving accross the barrier is of the order of the thermal energy $k_B T$. For a barrier consisting of an inverted parabola with a curvature determined by $\omega_S^B$, this condition yields $\Gamma/|\omega_S^B|\approx k_BT/U_b$, where $U_b$ is the energy barrier~\cite{haenggi90}. Experimentally, we measure the energy barriers $U_A\approx 4 k_B T$ and $U_C\approx 5 k_B T$ as well as  $|\omega_S^B|/2\pi\approx 51$~kHz.  Consequently, because $\Gamma/ P_\mathrm{gas} \approx 51$~Hz/Pa, we expect the turnover to occur at a gas pressure in the range $P_\mathrm{gas}\approx 1200$ to $1600$~Pa.  
In addition, we estimate the minimum energy path of the measured potential by using a steepest gradient algorithm (Fig.~\ref{fig:Pot}\textbf{c}). Following the minimum energy path we then evaluate the particle’s action $S$ over one oscillation period.
The action $S$ will be important later when we discuss the theoretical derivation of the particle's jumping rates.

Besides the reconstruction of the optical potential, we use the time traces of the particle position to determine the jumping rates between the wells. 
Figures~\ref{fig:Rates}\textbf{a-c} show typical time traces, projected along the $y$-axis (double-well axis), recorded for different gas pressures. The time traces clearly show the expected bistable behaviour. Moreover, the jumping rates change with gas pressure as expected from Kramers' rate theory. 
To confirm this observation we analyse the kinetics in greater detail.  
In the underdamped regime, the energy lost by the particle during one oscillation period is small, and hence once the particle has crossed the energy barrier it can recross it multiple times, before being retrapped in one of the two wells. 
Conversely, in the overdamped regime, the particle looses energy quickly and recrossing occurs only if the particle diffuses near the saddle point.
To eliminate the effect of correlated recrossings, we derive the jumping rate $R$ from the time autocorrelation of the binary population function $h_A(t)$, which is unity if the $y$ component of the particle position at time $t$ is negative ($y(t)<0$), {\it i.e.} if the particle is in well $A$, and equal to zero otherwise (see Supplementary Section~C). For kinetics governed by a rate equation, the time autocorrelation function $\langle \delta h_A(0) \delta h_A(t) \rangle $ is expected to exhibit an exponential decay at long times after an initial transient behavior caused by correlated recrossings~\cite{chandler78, dellago09}. Here $\delta h_A = h_A-\langle h_A\rangle$ is the deviation of $h_A$ from its long time average $\langle h_A\rangle$. By fitting the exponential decay $e^{-Rt}$ to $\langle \delta h_A(0) \delta h_A(t) \rangle$ we determine the jumping rate constant $R$ (Fig.~\ref{fig:Rates}\textbf{d}), which equals the sum of the rate constants for the forward and backward processes, $R = R_{AC}+  R_{CA}$.  
 Note that $R$ is the relaxation rate of a non-equilibrium population in the wells towards equilibrium. Besides removing correlated recrossings, this approach provides rates that are independent of the exact choice of the barrier separating the two stable wells $A$ and $C$.

\begin{figure}
    \begin{center}
        \includegraphics[width=.45\textwidth]{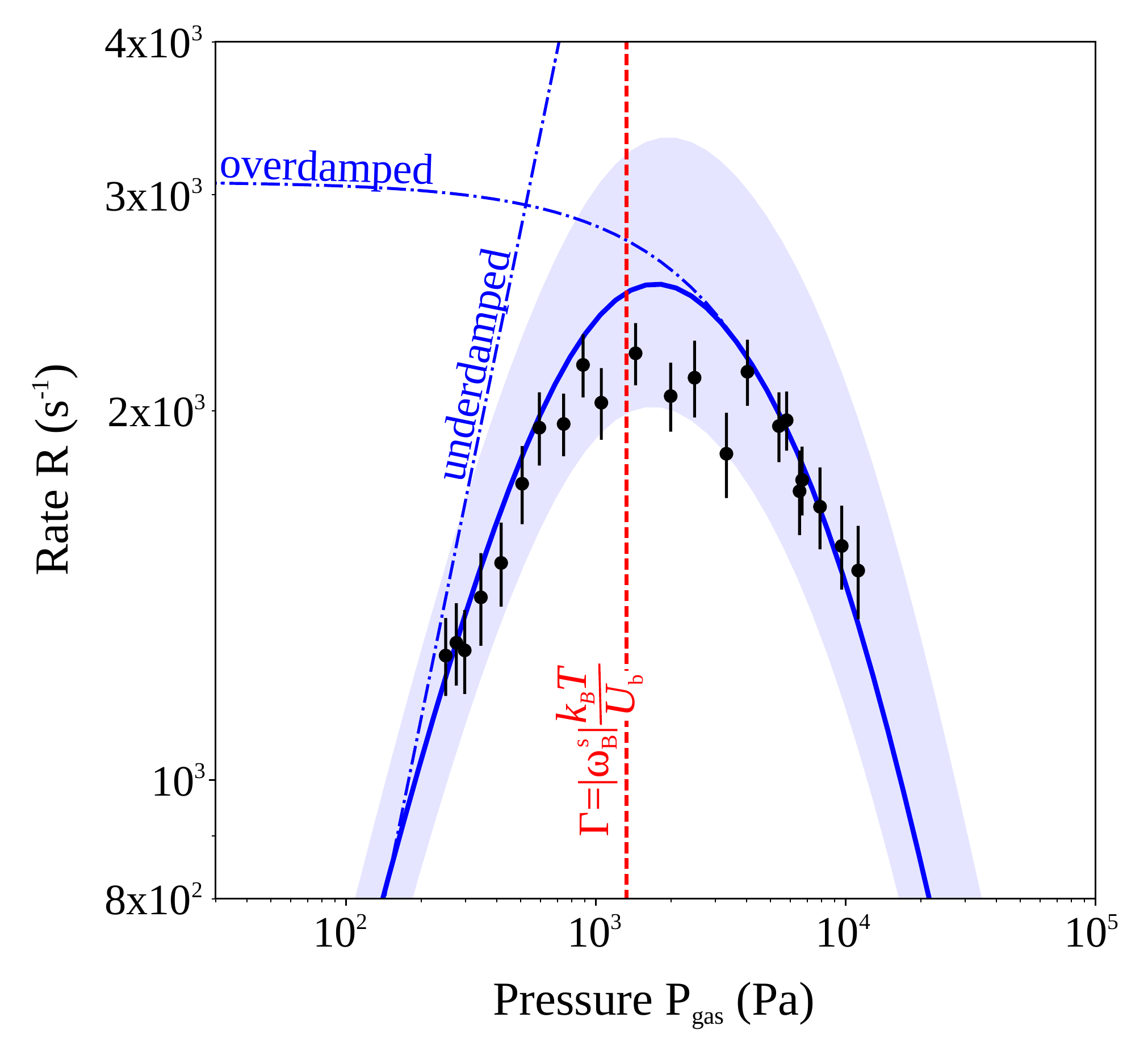}
    \end{center}
    \caption{Experimentally measured jumping rate $R$ as a function of gas pressure $P_\mathrm{gas}$ (black dots) compared with the analytical model of eq.~(\ref{eq:RateKT}) (solid blue line). The error bars indicate the standard deviation obtained by measuring the autocorrelation over 100 different time traces of 0.5 s. The blue shaded area indicates the uncertainty of the theoretical model due to parameter variations. The red line highlights the condition $|\omega_B^S|/\Gamma=k_BT/U_b$, where $U_b$ is an effective barrier height, corresponding to an average over the barrier heights measured for both trap $A$ and $C$. Dash-dotted lines highlight the two limiting cases of low and high damping (see Methods).
    \label{fig:KT}}
\end{figure}

To observe the Kramers turnover we record the dynamics of the particle over a wide range of pressures and hence dampings $\Gamma$. 
The chamber pressure  $P_\mathrm{gas}$ is lowered to 200 Pa ($\Gamma\approx 10$ kHz $\ll |\omega_S^B|k_B T/U_b$), and increased stepwise to $2\times 10^4$ Pa ($\Gamma\approx 1$~MHz $\gg |\omega_S^B|k_B T/U_b$). At each pressure step, the potential and the jumping rate are computed from time traces of the particle motion, as discussed previously.   
The measured rates are shown in figure~\ref{fig:KT}. The particle jumping rate clearly exhibits a maximum, near $\Gamma = |\omega_S^B| k_B T/U_b$. \\[-2.5ex]

The experimentally determined optical potential allows us to quantitatively compare our measurements  with theoretical models that have been developed during the last decades~\cite{haenggi90,melnikov91,pollak13}.
In the overdamped regime, multidimensional rate theory, which has been studied intensively theoretically~\cite{hershkovitz97}  and verified experimentally~\cite{han89,mccann99}, yields
\begin{equation}
    R_{AC}^\mathrm{HD} = \frac{1}{2\pi} \prod_{i\in\{x,y,z\}}  \frac{\omega_i^A}{|\omega_i^B|} \left[ \sqrt{|\omega_B^S|^2+\frac{\Gamma^2}{4}}-\frac{\Gamma}{2}\right] e^{-\frac{U_A}{k_BT}} \ ,
    \label{eq:R_HD}
\end{equation}
for the transition from well $A$ to $C$. The reverse rate $R_{CA}$, for the transition from well $C$ to $A$ is obtained by swapping the indices $A$ and $C$. 
For a given barrier height, the rate constant decreases with increasing damping and becomes inversely proportional to $\Gamma$ in the limit of high friction due to slow diffusion of the system across the barrier.

In the underdamped regime, on the other hand, the rate limiting factor is the slow transfer of energy between system and bath, leading to rate proportional to $\Gamma$. While the jumping rates in the two limiting cases where already derived by Kramers, a full analytical theory bridging the underdamped and overdamped regimes was obtained only much later~\cite{melnikov91}. To cross-over from the overdamped to the underdamped case, one introduces the depopulation factor $\Upsilon(\delta) = \exp \left[\frac{1}{\pi} I(\delta)  \right]$,  where 
\begin{equation}
    I(\delta) = \int_{-\infty}^\infty\ln\left\{ 1-\exp\left[-\frac{\delta}{k_B T} (x^2 +\frac{1}{4}) \right] \right\}\frac{\mathrm d x}{x^2 +\frac{1}{4}}  \ ,
    \label{eq:I}
\end{equation}
 $\delta$ is the energy loss parameter and the integration is carried out over the positive part of the real axis~\cite{haenggi90,melnikov91}. 
The estimation of this energy loss parameter is one of the major challenges in the context of Kramers' turnover theories~\cite{pollak13}. 
However, in our experimental situation, where the  friction is memory free, this energy loss is well approximated by $\delta = \Gamma S$, where $S=S_A$ [resp. $S_C$] is the particle action over one oscillation
for the particle in well $A$ [resp. $C$]~\cite{melnikov91}. The action is then measured along the minimum energy path of the potential, shown in Fig.~\ref{fig:Pot}\textbf{b} and \textbf{c},  from $A$ to $B$ [resp. $C$ to $B$] (See Methods).
In the case of a double-well potential, due to the recrossing dynamics, the probability of the particle to be  retrapped in trap $A$ is different from that in trap $C$. 
The correct rate expression for the rate dynamics, for any damping, requires to account for these different probabilities. The transition rate from $A$ to $C$ is thus obtained by multiplying the high damping transition rate $R_{AC}^\mathrm{HD}$ (eq.~\ref{eq:R_HD}), by the factor $\Upsilon(\Gamma S_A) \Upsilon(\Gamma S_C)/\Upsilon(\Gamma S_A+\Gamma S_C)$~\cite{haenggi90,melnikov91}. 
Finally, our rate estimates, based on measuring the autocorrelation of the particle's time traces, yields the sum of the rates  $A\rightarrow C$ and $C\rightarrow A$. 
Therefore, the expected theoretical rate is
\begin{equation}
    R (\Gamma) = \frac{\Upsilon(\Gamma S_A) \Upsilon(\Gamma S_C)}{\Upsilon(\Gamma S_A+\Gamma S_C)}\left[ R_{AC}^\mathrm{HD}+ R_{CA}^\mathrm{HD}\right]\, ,
    \label{eq:RateKT}
\end{equation}
which is valid in the entire range from low to high friction.
%

To mitigate the influence of small changes of experimental parameters (e.g. due to drift), the optical potential is measured for each pressure step and the experimental parameters used in the rate analysis (eq.~(\ref{eq:RateKT})) are  determined independently for each pressure~(see Methods and Supplementary Section~D).  
To compare the analytical model with our measurements, we average the parameters obtained for different pressures and use them in eq.~(\ref{eq:RateKT}). The resulting rate is presented in figure~\ref{fig:KT} (solid blue line). We then calculate the uncertainties of the model by evaluating the ensemble of the measured parameters over the whole set of pressures. These uncertainties are pictured as a blue shaded area. 
Our experimental data always fall within the uncertainty and agree to the prediction obtained with averaged parameters to within less than 20\%. This small mismatch between theory and experiment is caused by uncertainties in the determination of the exact shape of the potential as well as drifts associated with the measurement beam,  which are more pronounced at higher pressures (see Supplementary Section~F for details). Nevertheless, our results give a strong support to the analytical model presented in Eq.~(\ref{eq:RateKT}). We anticipate that improvements in the stability of the setup, especially by having a precise pressure and temperature control in addition to improved detection will further enhance the agreement.

The laser-trapped nanoparticle serves as a statistical simulator that allows us to study rate theories in complex systems.
The particle dynamics can be studied in time dependent arbitrary three-dimensional energy landscapes that can be created by optical point-spread-function engineering~\cite{bowman13}. 
We can also study non-ohmic friction by applying correlated noise to the trapping beams. 
Thus, our system provides an experimental testboard for a large variety of physical models.
In addition, theoretically challenging non-equilibrium dynamics can be easily addressed with our statistical simulator, either by applying noise to the trapping potential~\cite{dykman05}, or by preparation of non-thermal states~\cite{gieseler14}. This enables the study of different noise-induced phenomena, such as stochastic resonance~\cite{gammaitoni98,ricci17}, in any damping regime. 
Finally, given the recent advances toward quantum ground state cooling of vacuum-trapped particles~\cite{kiesel13,fonseca16,jain16}, we expect that our system will provide a way to shed light on rate theories at the classical to quantum transition.
This ability to test the applicability and the potential break-down of escape rate models is of crucial interest to understand and interpret experiments.

To summarize, using an optically trapped nanoparticle in vacuum we directly measured  Kramers' turnover. The experimental approach provides quantitative data needed to test multidimensional  rate theories  and explore  parameters that influence the escape dynamics from single potentials and hopping between stable wells. Due to the fine control of the system parameters, our experimental simulator can be  used to study switching dynamics  and validate  rate models in a wide range of system-bath couplings, shining light on problems ranging from protein folding~\cite{chung15} to diffusion in solids~\cite{guantes03}. Also,
the experimental platform is well suited for studying  open questions in non-equilibrium and non-Markovian statistical physics, in particular in situations where  Brownian dynamics is influenced by different damping regimes.  Such studies will benefit the development of optimal protocols for information to energy conversion~\cite{toyabe10,berut12} and stochastic heat engines~\cite{dechant15}.  

\begin{acknowledgments}
This research was supported by Swiss National Science Foundation (no. 200021L\_169319) in cooperation with the Austrian Science Fund (no. I 3163), ERC-QMES (no. 338763), CoG ERC-QnanoMECA (no. 64790), Fundaci{\'o} Privada CELLEX and the severo Ochoa program. LR acknowledges support by an ETH -- Marie Curie Cofund Fellowship.
The authors thank M. Frimmer, V. Jain, E. Hebestreit, C. Moritz, P. Mestres, E. Pollak and  P. Bharadwaj for discussions and experimental support.
\end{acknowledgments}

\section*{Methods}

\subsection*{Experimental setup}
The optical double well trap is generated using two cross-polarized, and frequency shifted (40 MHz) beams from a continuous-wave infrared laser ($\lambda=1064$~nm, Coherent Mephisto). Before entering the chamber, the two orthogonal polarizations are recombined using a polarizing beam splitter (PBS), and focused by a N.A.=0.8 objective (Nikon). The relative position of the two traps can be changed using a steering mirror (SM, Newport Fast Steering Mirror). \\
The particle position is recorded using an additional, weakly focused, green laser ($\lambda = 532$~nm, Coherent) and three balanced detectors (Newport, 1817-FC). The signal is filtered (low pass filter at 1 MHz) before recording on a fast data acquisition card (Gage,  Oscar CS4342). The 3D particle trajectory is reconstructed from the measured signal using a calibration function, determined independently. Additional details on the experimental setup and the measurement calibration can be found in the Supplementary Sections~A and~B.   

\subsection*{Determination of physical parameters}
~\\
\noindent \underline{1. Positions of the saddle  and  stable points :}
The optical potential is derived from the particle distribution function $\rho(\mathbf r)$ obtained by histogramming the particle position time trace. Thus, the optical potential $U$ at each point  $\mathbf r$ follows from 
\begin{equation}
    U(\mathbf r) = -k_B T \ln\left[\rho(\mathbf r) \right] \; ,
    \label{eq:U}
\end{equation}
where $k_B$ is the Boltzmann constant and $T =300$~K is the temperature of the gas. By differentiation we obtain the effective force field
   $\mathbf F(\mathbf r) = -\mathrm{\bf grad}\  U(\mathbf r)$.
 The force vanishes at the two stable points $A$ and $C$, and at the saddle point $B$.

\noindent \underline{2. Energy barriers and potential curvature :}
Near the saddle point and the stable points, the potential energy can be approximated by a harmonic function, 
\begin{equation}
    U_\mathrm{HO}(\textbf r) = U_\alpha +\frac{1}{2} \sum_{i,j}\Lambda^\alpha_{i,j} (r_i-r_i^\alpha)  (r_j-r_j^\alpha) \; ,
    \label{eq:Uho}
\end{equation}
where $\alpha=A, B,$ or $C$, $\textbf r^\alpha$ is the spatial position of point $\alpha$, $U_\alpha$ is the potential energy at point $\alpha$, and $i,j\in \{x,y,z\}$.

\noindent A local fit of the  measured  potential $U_\mathrm{HO}$ yields the matrix $(\Lambda^\alpha_{i,j})$ for each of the three spatial points.
The  three eigenvalues  $\lambda^\alpha_{k}$ of each matrix correspond to the curvatures along the three local normal axes $k\in\{x',y',z'\}$ of the potential, $\omega_k^\alpha=\sqrt{{\lambda^\alpha_{k}}/{m}}$, where $m$ is the mass of the particle, and the primed coordinates denote the local coordinate system. Alternatively, the full potential can be fitted by
\begin{equation}
    \Pi_\mathrm{fit}(x,y,z)=\sum_{i,j,k}\mu_{i,j,k}x^iy^jz^k
    \label{eq:globFit}
\end{equation}
to derive the curvature in any point of the potential. Good convergence is typically obtained for a polynomial at order $\sim10$. The results of the two different fitting procedures agree within a few percent.\\

\noindent \underline{3. Particle action :}
The particle action over one oscillation is approximated by the action over the minimum energy path,  given by 
\begin{equation}
    S_{i} = 4\int_{r_{i}}^{r_B} \sqrt{2m[U_B-U(r)]}\mathrm d r\ ,
    \label{eq:S_A}
\end{equation}
{where $i = A,C$}.\\
The integration is carried out along the minimum energy path, which is the path between the two stables points $A$ and $C$ that minimizes the energy. We determine the  minimum energy path using a steepest gradient algorithm. This path is shown in Fig.~\ref{fig:Pot}\textbf{c}. As a guide for the eye, the path in this figure has been continued beyond $A$ and $C$, although there is no minimum energy path outside of the $A$ to $C$ range. Finally, note that the factor $4$ in Eq.~(\ref{eq:S_A}) originates from measuring only one fourth of the full oscillation period.

\noindent The evolution of recorded parameters as a function of pressure is presented in the Supplementary Information section (Fig~S7).\\

\subsection*{Low and high damping limiting cases.}
~\\
\noindent Figure~\ref{fig:KT} presents the theoretical limiting cases, corresponding to the low and the high damping regimes (blue dash-dotted lines). The high damping rates have been computed using eq~(\ref{eq:R_HD}), while the low damping rates correspond to the equation~\cite{melnikov91}:
\begin{equation}
    R^\text{LD}=\dfrac{S_A S_C}{S_A+S_C}\dfrac{\Gamma}{2\pi k_B T}\left( \omega_{y'}^A e^{-\frac{U_A}{k_BT}}+ \omega_{y'}^C e^{-\frac{U_C}{k_BT}} \right)\ .
    \label{eq:R_LD}
\end{equation}
Note that this equation is an approximation for the low damping regime in one dimension that should be a good approximation for our three-dimensions case since the three normal modes are only weakly coupled~\cite{hershkovitz97}.


\setcounter{figure}{0}
\renewcommand{\thefigure}{S\arabic{figure}}

\newpage
\begin{widetext}
\section*{Supplementary Information}

\subsection{Experimental Setup}
An optical double well trap is generated using two cross-polarized beams from a continuous-wave infrared laser ($\lambda=1064$~nm), split by a polarizing beam splitter (PBS) (see Fig.~\ref{fig:setup}). One polarization is frequency shifted by 40 MHz with an acousto-optic modulator (AOM) to suppress effects related to interferences of the two trapping beams. Before entering the chamber, the two orthogonal polarizations are recombined using a second PBS, and focused by a N.A.=0.8 objective. The transverse position of the focus depends on the beam angle at the objective back-focal-plane. The relative angle of the two trapping beams can be easily changed using a telecentric $4f$ setup combined with a steering mirror (SM, Newport Fast Steering Mirror). In addition, the optical power of each trapping beam can be adjusted using half-wave plates (HWP).  
Therefore, by changing the angle and adjusting the power we can tune the symmetry and the depth of  the optical double-well potential. \\

\begin{figure}[!h]
    \begin{center}
        \includegraphics[width=\textwidth]{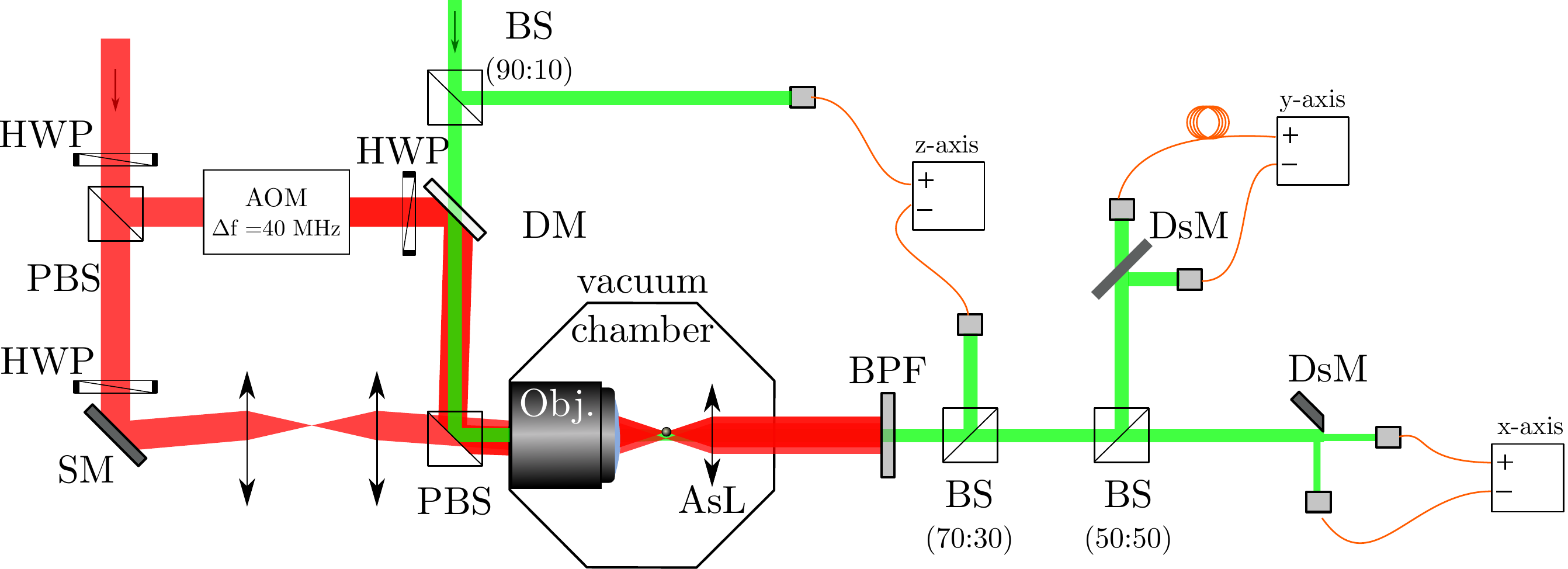}
    \end{center}
{\it     \caption{Experimental setup.  \label{fig:setup}}}
    \vspace{1em}
\end{figure}
%
An additional green laser ($\lambda = 532$~nm) is used for measurement. It is combined with the trapping beams using a dichroic mirror (DM).  The beam size of the measurement laser at the objective back-aperture is chosen such that the effective NA for this beam is roughly $0.2$. This helps to reduce the nonlinearities in the position measurement over the whole   optical double trap. 
The measurement beam is then recollimated with an aspheric lens (AsL), and isolated from the trapping beam using a 532~nm band pass filter (BPF).\\

The detection scheme is based on a common path interferometer and is described in Ref.~\cite{gieseler12SI}. In short, the motion of the particle is imprinted in the phase of the re-scattered light, which interferes with the non-scattered transmitted light beam. The phase modulation is thus converted into an amplitude modulation that is directly recorded using fiber coupled 80 MHz balanced detectors (1807-FC, Newport).
For a particle close to the focus of the measurement beam, recording the raw signal provides a signal proportional to the particle's $z$-motion. Similarly,  by splitting the beam vertically [or horizontally] with a D-shape mirror (DsM), a signal proportional to the $x$ [or $y$] motion is obtained.
As discussed in the next section, for large displacement of the particle from the optical axis, non linearities have to be taken into account in the measurement scheme.

\subsection{Detection scheme calibration}

\begin{figure}[!ht]
    \begin{center}
        \includegraphics[width=.7\textwidth]{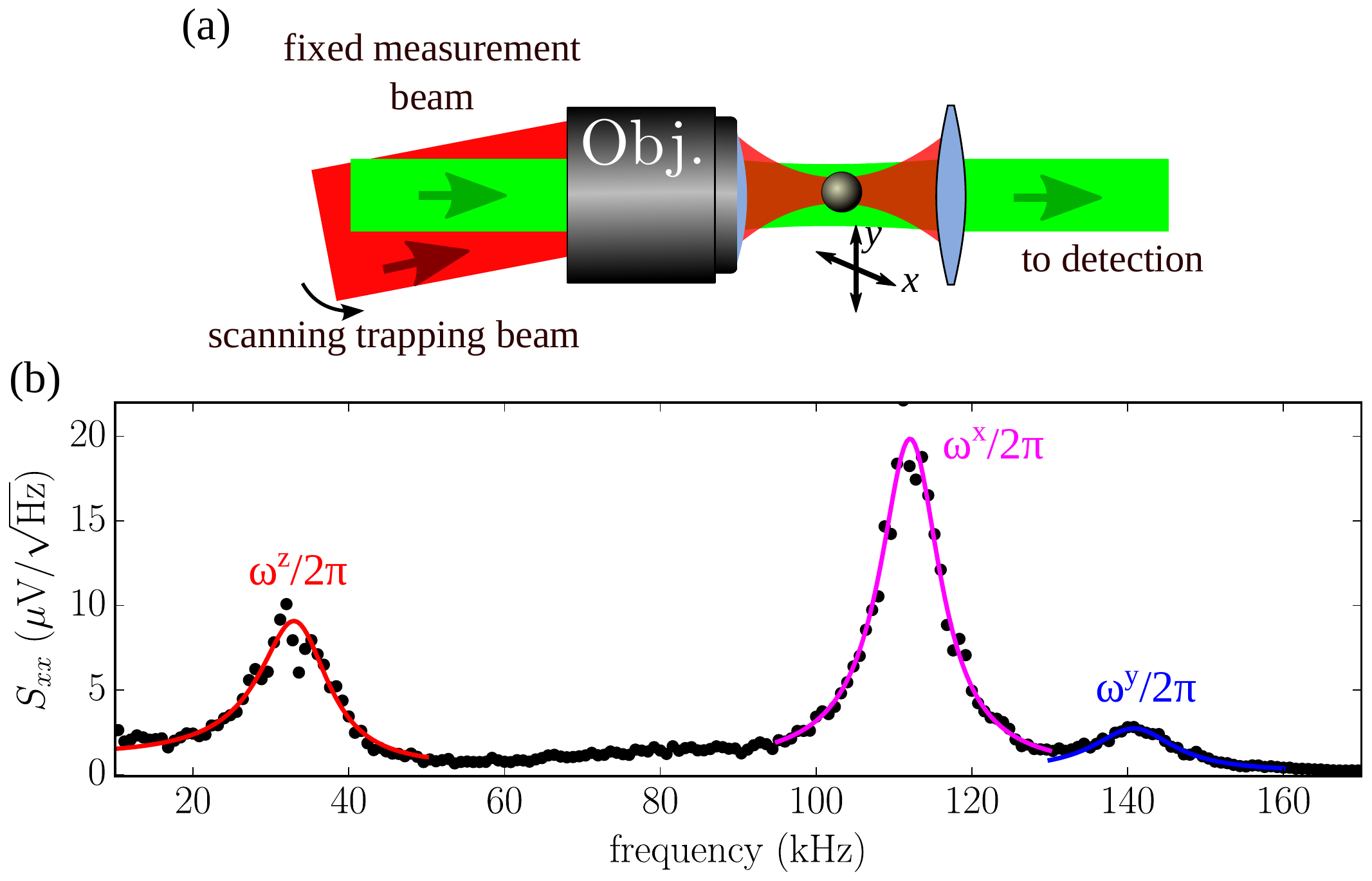}
    \end{center}
    \vspace{-1em}
   {\it  \caption{Measurement scheme calibration. (a) Experimental setup used for the measurement scheme calibration. (b) Power spectral density of the $x-$motion recorded for a particle trapped away from the optical axis. The PSD shows peaks at characteristics frequencies corresponding to the three eigenfrequencies of the trap $\omega^z/2\pi$, $\omega^x/2\pi$ and  $\omega^y/2\pi$. These three peaks are fitted independently (solid lines).      \label{fig:Sxx_11x11}}}
    \vspace{1em}
\end{figure}
A voltage is recorded on the three detectors. To reconstruct the dynamics of the particle the recorded voltage has to be expressed as a distance in nanometers through a calibration factor $\mathcal C_\mathrm{calib}$. Recent experiments involving levitated particles in vacuum work in a regime where the recorded signal is simply linear with the  particle position and, as a consequence, the calibration factor  $\mathcal C_\mathrm{calib}$ is constant. In the present work, the particle displacement is large and the detection scheme is no longer linear. 
To account for these nonlinearities we introduce the calibration functions $\mathcal C_\mathrm{calib}^q(\vec r)$, where $q\in\{x,y,z\}$ is the considered axis, and $\vec r$ the position of the particle.\\

\begin{figure}[!ht]
    \begin{center}
        \includegraphics[width=.6\textwidth]{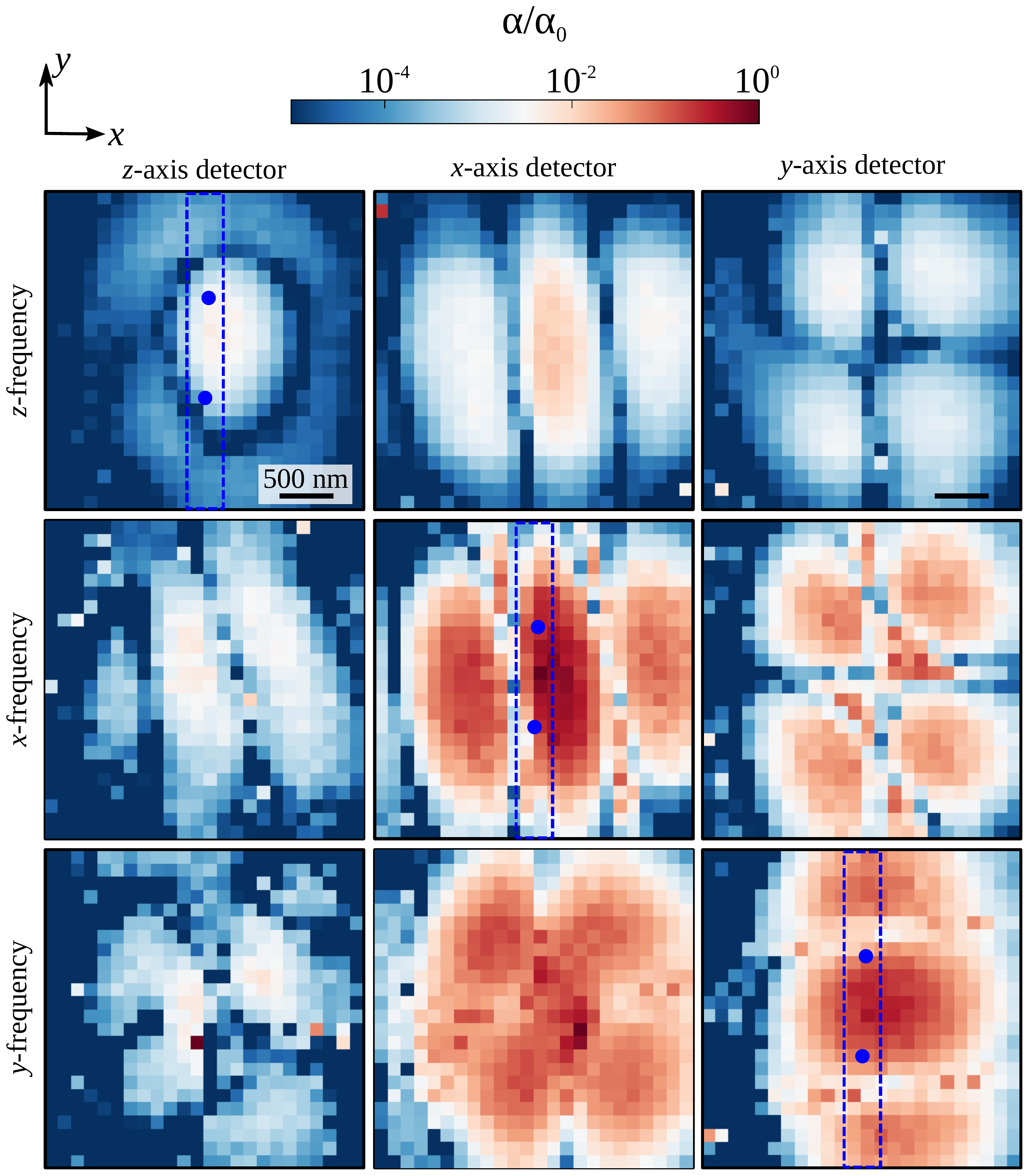}
    \end{center}
    \vspace{-1em}
   {\it  \caption{Values of the amplitude $\alpha$ of the PSDs obtained from fits according Eq.~(\ref{eq:Sfit}), normalized by its maximum value, of the $x$, $y$ $z$ motion (lines) on the $x$, $y$ and $z$ detector (columns), as a function of the particle's position in the focal plane.  Off diagonal figures highlight the measurement axis cross-coupling. The blue dots correspond to the position of trap $A$ and trap $C$ of the potential described in the main text. The dotted areas correspond to the areas used for the calibration of our measurement in the main text.
    \label{fig:SensXY}}}
    \vspace{1em}
\end{figure}

To experimentally determine the  $\mathcal C_\mathrm{calib}^q$ functions, a particle is trapped using a single trapping beam.
The particle is then scanned through the focal plane of the measurement beam by displacing the trap with a fast steering mirror (Fig.~\ref{fig:Sxx_11x11}-(a)). 
At each point in the focal plane the particle dynamics is recorded for 2~s using a fixed measurement beam. Our telecentric measurement scheme allow us to keep the axes of the particle's trapping potential aligned with the axes of the measurement beam.
 The power spectral density (PSD) is computed from the recorded time traces for the 3 measurement axes $x$, $y$ and $z$. Each of the PSDs can features 3 peaks at the three eigenfrequencies $\omega_x$, $\omega_y$, and $\omega_z$, which correspond to the three trap axes. The presence of three frequencies is due to cross coupling between the different measurement axes. Indeed, under certain circumstances the $x$-detector may, for example, detect a signal coming from  the motion along to $y$ and/or $z$ axis (Fig.~\ref{fig:Sxx_11x11}-(b)).
 To completely characterise the detection nonlinearities and associated axes cross-coupling, each of the three peaks is fitted by the function
\begin{equation}
    S_{qq}^\mathrm{fit}(\omega) = \sum_{i\in\{x,y,z\}} \frac{\alpha_{qi}}{(\omega_i^2-\omega^2)^2+\omega^2\Gamma_i^2}+\beta \, ,
    \label{eq:Sfit}
\end{equation}
where $\alpha_{qi}, \beta, \Gamma_i$ and $\omega_i$ are adjustable parameters, and $q$ is the measurement axis.
As an example, a PSD $S_{xx}$ recorded for the particle away from the measurement beam's centre is shown in Fig.~\ref{fig:Sxx_11x11}. The measurement is overlaid with analytical fitting functions according Eq.~(\ref{eq:Sfit}).\\

The amplitudes $\alpha_{qi}$ of the three peaks present in the PSD provide information about a detector's sensitivity to the particle motion, both along the axis of interest and along the two orthogonal axes (cross-coupling). These amplitudes, normalized by the overall maximum amplitude $\alpha_0$, for the three trap frequencies and for each detector are shown in Fig.~\ref{fig:SensXY}.
As expected, we observe that the detection sensitivity is constant only near the centre of the measurement beam. For larger displacements the sensitivity drops and cross-couplings come into play. 
Such a behaviour is theoretically expected and has already previously been used to extend the spatial range of particle dynamics studies~\cite{perrone08SI}. \\

In our experiment, the two potential wells are aligned along the $y$-axis. Consequently, large displacements are recorded along the $y$-axis, whereas displacements along the $x$ and $z$-axes remain small (blue dotted rectangular area in figure~\ref{fig:SensXY}) we assume that cross-couplings can be neglected and that the detection sensitivity depends only on $y$-axis. 
By virtue of  the equipartition theorem, the integrated area under the power spectral density is proportional to the bath temperature $k_B T$~\cite{gieseler12SI}. Therefore, from the fit of the PSD we extract the calibration function for each axis, 
$$\mathcal C_\mathrm{calib}^q(y)= \sqrt{\frac{\alpha^q(y)\pi m}{\Gamma k_B T}}\ ,$$
which  converts the measured voltage on the detector to particle displacement in meters. To reduce noise, this calibration function is integrated along the $x$-axis in the blue dotted rectangular area of figure~\ref{fig:SensXY}.
The obtained calibration functions for the three different axes are shown in figure~\ref{fig:CalibXYZ}. These functions are fitted by a 10th order polynomial (red solid line) that is used to calibrate the particle position from the recorded time traces. 
From these calibration functions we estimate the sensitivity of our measurement. Typical noise on the detectors is on the order of 300~$\mu$V. Given that the conversion factor for  $x$-and $y$-axes  is better than 50~nm/mV, we deduce that the position is known with a precision better than 15~nm. For the $z$-axis, the sensitivity is almost ten times worse due to the use of a weakly focused measurement beam. However, errors in the $z$-axis position only weakly affect the theoretical rate estimate,  since the $z$-axis is orthogonal to the trapping axis.\\
\begin{figure}[!t]
    \begin{center}
        \includegraphics[width=\textwidth]{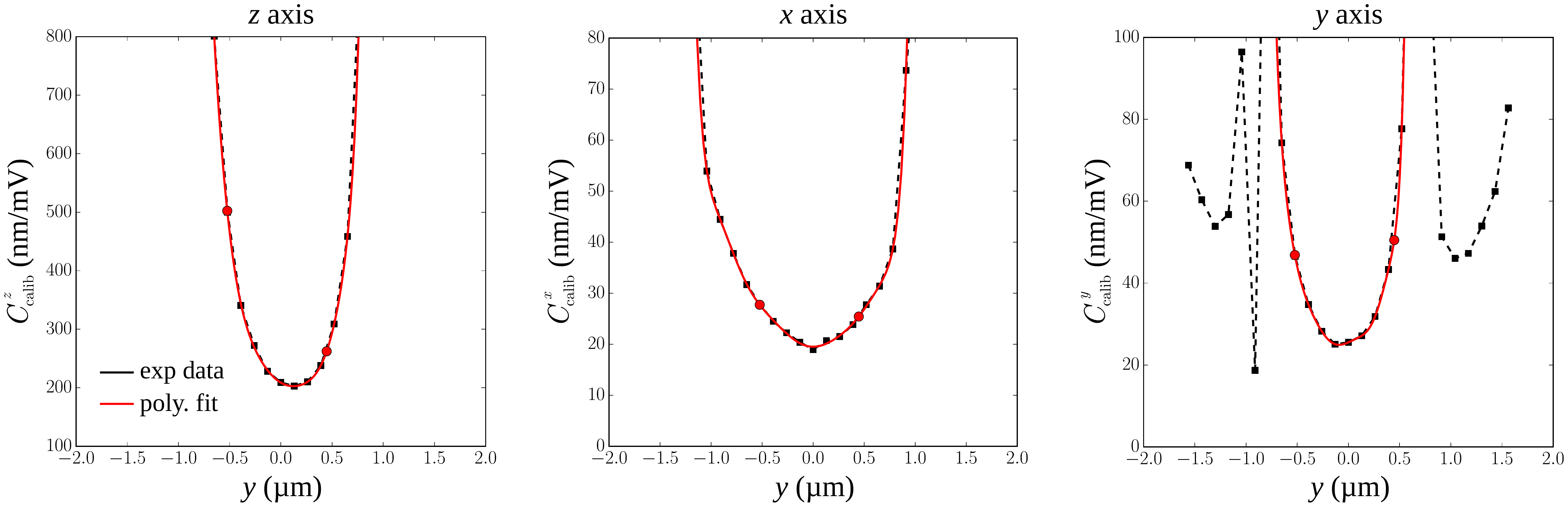}
    \end{center}
    \vspace{-1em}
   {\it  \caption{Particle displacement calibration. Calibration as a function of the particle position along the $y$-axis, obtained by averaging along $x$ over the blue dotted area of figure~\ref{fig:SensXY}-(b). The experimental data (black squares) are fitted by a 10th order polynomial function (red line), which is used for calibration of the particle dynamics. 
       \label{fig:CalibXYZ}}}
    \vspace{1em}
\end{figure}

Measuring the PSD of the particle dynamics allows us to also retrieve information on the gas damping $\Gamma$. The particle, while trapped in the centre of the measurement beam, is recorded at different gas pressures and the PSD is computed (Fig.~\ref{fig:calib}-(a)). 
\begin{figure}[!ht]
    \begin{center}
        \includegraphics[width=.76\textwidth]{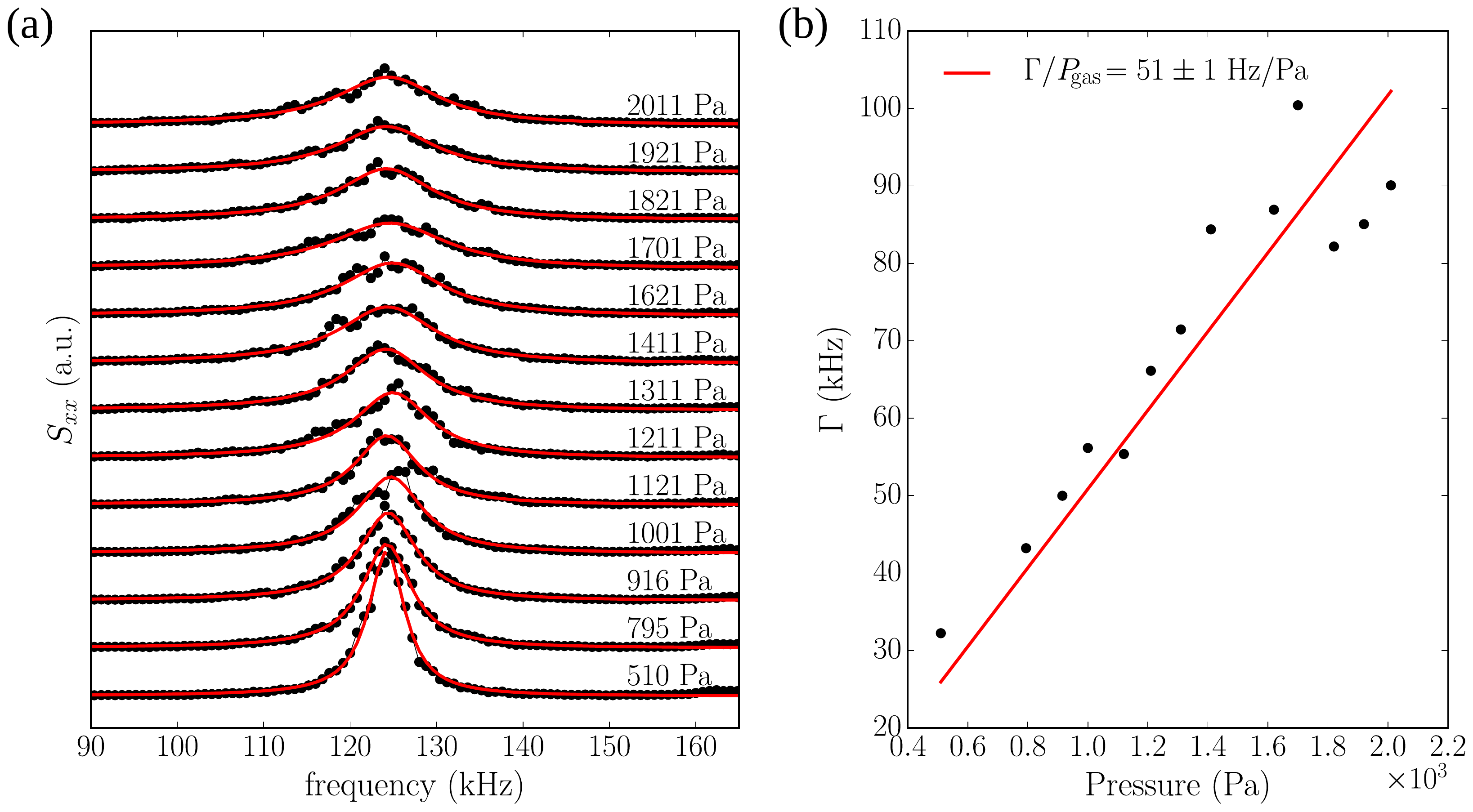}
    \end{center}
    \vspace{-1em}
   {\it  \caption{Damping measurement. (a) Power spectral density of the $x-$motion recorded at different gas pressures. A fit (red solid line) allows us to retrieve the calibration factor as well as the damping $\Gamma$. (b) Experimentally measured damping $\Gamma$ from the fit in (a), as a function of gas pressure $P_\mathrm{gas}$. A linear fit gives the coefficient that is used in the main text for the conversion from  $P_\mathrm{gas}$ to $\Gamma$ (red solid line).
   \label{fig:calib}}}
    \vspace{1em}
\end{figure}
From the fits we retrieve the value of the gas damping $\Gamma$ and verify its linear relationship with the pressure $P_\mathrm{gas}$ (Fig.~\ref{fig:calib}-(b)). As expected from kinetic theory for a 68~nm radius particle~\cite{gieseler12SI,beresnev90SI}, we experimentally measure  $\Gamma/P_\mathrm{gas} \approx 51$~Hz/Pa.\\

\subsection{Rate Measurements}
To determine the particle jumping rate we split the space into two sub-spaces $\mathcal A$ and $\mathcal C$. These two sub-spaces are separated by a dividing surface, which sets the boundary between the two traps $A$ and $C$. Transition rates are extracted from the time autocorrelation function $\langle \delta h_A(0)\delta h_A(t)\rangle$, where $h_A$ is the  binary population function defined as 
$$
h_A(t) = \left\{
    \begin{array}{ll}
        1 & \mbox{if  the particle is measured in subspace $\mathcal A$ at time } t \\
        0 & \mbox{if it is measured in $\mathcal C$  ,}
    \end{array}
\right.
$$
and $\delta h_A=h_A-\langle h_A \rangle$ is the fluctuation of $h_A$.\\

With this approach, the resulting rate constant is not affected by the exact position of the dividing surface.
Therefore, for simplicity, we choose the $x\!-\!z$ plane as a separating surface. 
Thus, the population binary function is obtained by assigning to each measurement point $\{x(t),y(t),z(t)\}$ of the particle motion,  the value 
$$
h_A(t) = \left\{
    \begin{array}{ll}
        1 & \mbox{if \ } y(t)<0 \\
        0 & \mbox{else.}
    \end{array}
\right.
$$
An example of the time evolution of the  population function $h_A$ is shown in figure~\ref{fig:hA}.
\begin{figure}[!htb]
    \begin{center}
        \includegraphics[width=.7\textwidth]{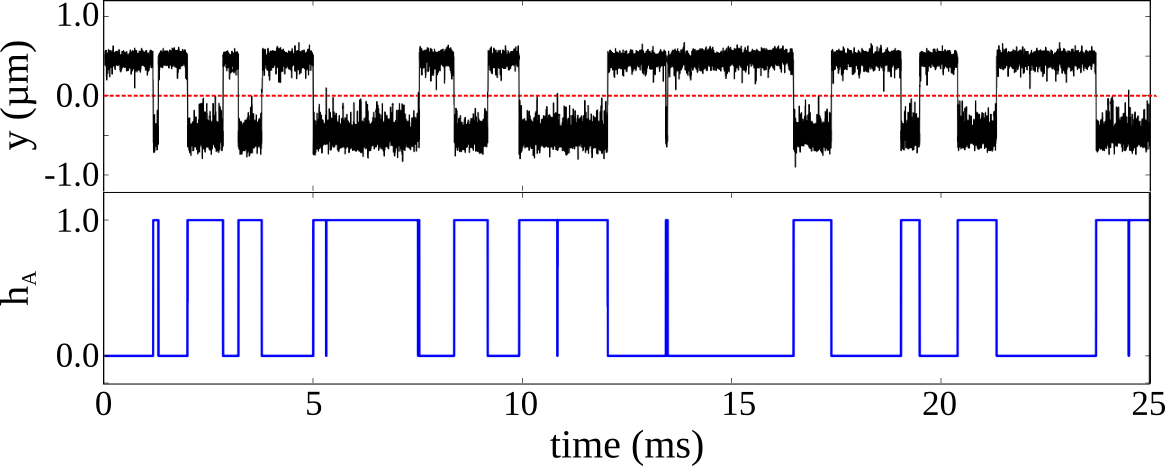}
    \end{center}
    \vspace{-1.5em}
{\it     \caption{The  binary population function $h_A(t)$ (blue line) is computed from the time trace projected along the axis $y$ (black line). The red dashed line indicates the position of the dividing surface.
    \label{fig:hA}}}
        \vspace{1em}
\end{figure}

The time autocorrelation  $\langle \delta h_A(0) \delta h_A(t)\rangle$ of a particle jumping between two traps is characterized, in general, by a transient for short times (due to correlated recrossings) followed 
by an exponential decay~\cite{dellago09SI,chandler78SI}. An exponential fit $e^{-Rt}$ of the long-time behaviour yields the jumping rate $R = R_\mathrm{AC}+R_\mathrm{CA}$. \\

\newpage
\subsection{Evolution of Physical Parameters\\[-2ex]}

As mentioned in the main text,  the optical potential is affected by slow drifts. Therefore, the parameters of the optical potential have to be re-evaluated for each setting of the vacuum pressure. Figure~\ref{fig:PotCar} shows the variations of normal mode frequencies and energy barriers as a function of gas pressure.
%
\begin{figure}[!h]
    \begin{center}
        \includegraphics[width=.45\textwidth]{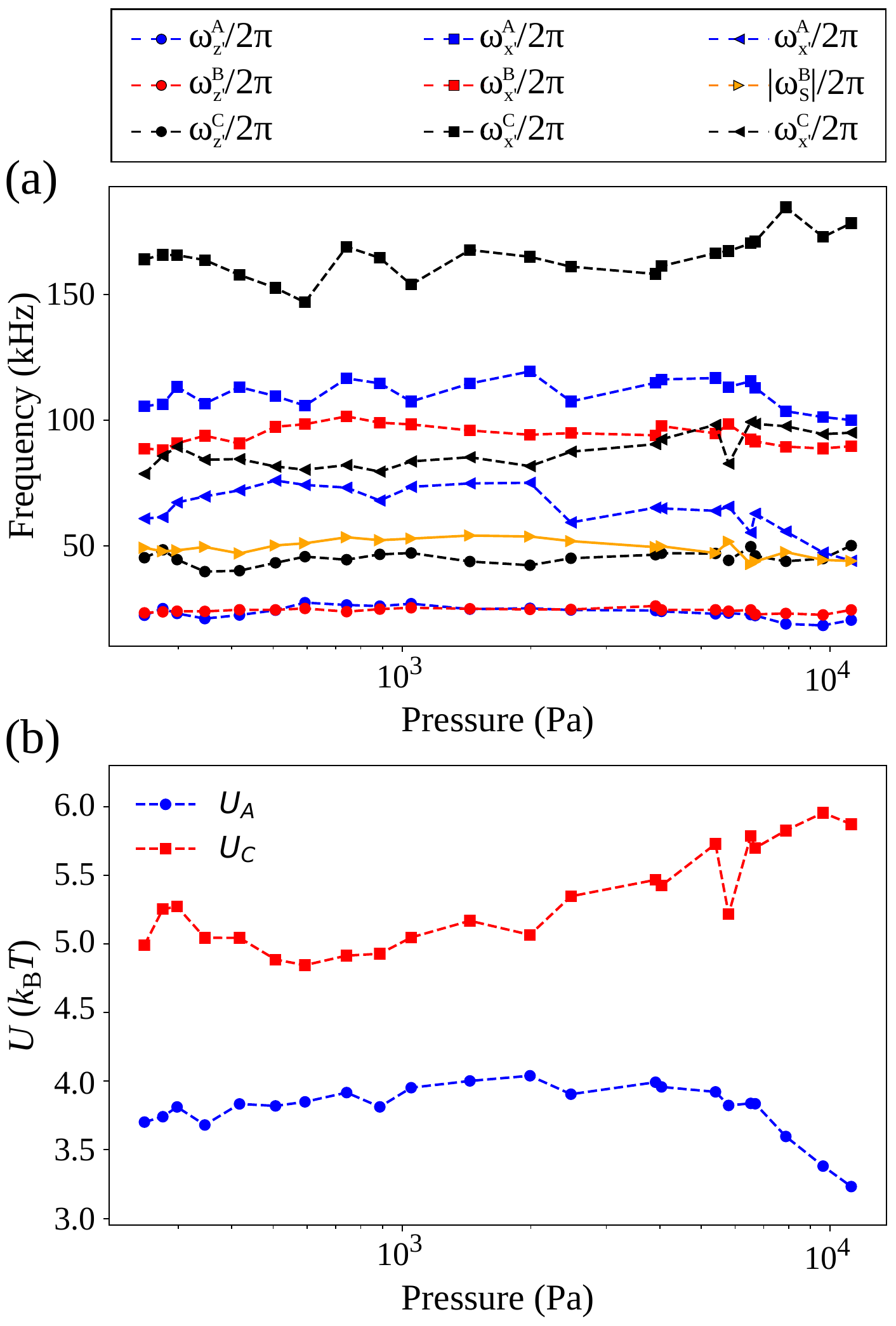}
    \end{center}
    \vspace{-1.5em}
{\it    \caption{Evolution of parameters as a function of gas pressure. (a)~Normal mode frequencies $\omega_i^A/2\pi$ (blue), $\omega_i^B/2\pi$ (black) and $\omega_i^C/2\pi$ (red). The saddle point frequencies $|\omega_S^B|/2\pi$ are shown in orange. (b)~Energy barriers $U_A$ (blue circles) and $U_C$ (red squares).
    \label{fig:PotCar}}}
\end{figure}
%

\subsection{Estimation of the damping rate}
The damping rate is central to calculating the transition rates. For a levitated particle the damping rate is proportional to the gas pressure inside the experimental chamber~\cite{gieseler12SI,beresnev90SI}. Therefore, in principle, we can directly measure the damping rate using our pressure gauge (Pirani gauge, Instrutech),  once the proportionality factor between pressure and damping factor is know.
As shown in the calibration subsection (Figure~\ref{fig:calib}), a fit to the PSD of the particle dynamics provides an estimate of the damping term, and is used to determine the relation between damping and pressure at pressure.
%
\begin{figure}[!h]
    \begin{center}
        \includegraphics[width=.5\textwidth]{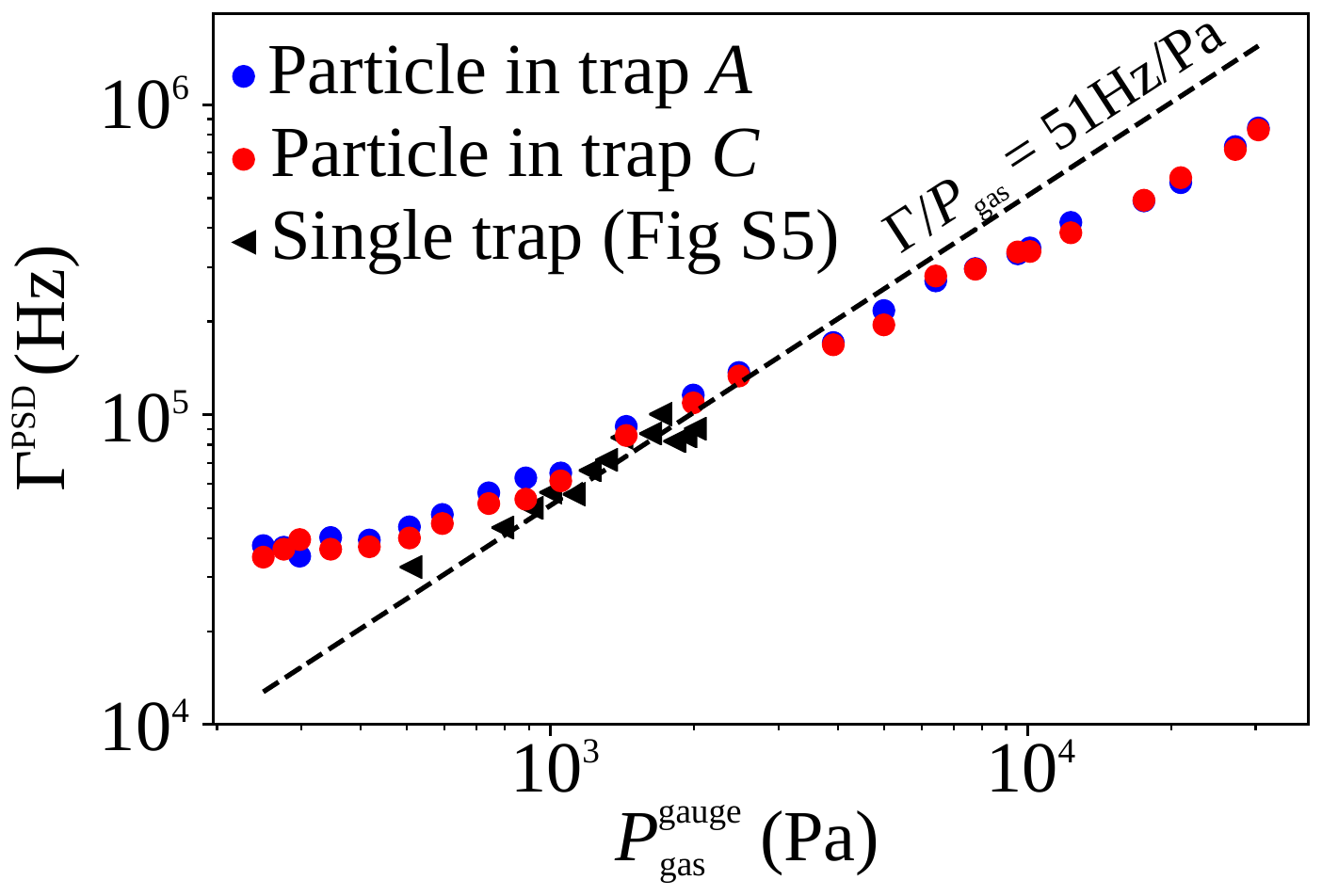}
    \end{center}
    \vspace{-1.5em}
    {\it    \caption{Width of the PSD $\Gamma^\text{PSD}$ for the motion of the particle along the $x$-axis, for the particle localised in trap $A$ (blue dots) and in trap $C$ (red dots), as a function of the pressure (as recorded by the gauge). The black dotted line corresponds to the theoretical relation between damping and gas pressure. Data taken from a single trap (Fig. S5) are shown for comparison (black triangles). 
    \label{fig:GamVsP}}}
\end{figure}
%
However, it is important to note that this approach is only valid when the non-linearities of the potential are weak compared to the damping,  i.e for pressures above $\approx 500$~Pa for a single beam trap~\cite{gieseler13SI}.

In the following we estimate the damping constant from the particle motion when the particle is trapped in a double well potential.
A double trap setup is intrinsically non-linear. To overcome this limitation we isolate the dynamics of the particle when localised in one of the traps. 
Using a threshold algorithm, we segment the timetrace of the particle motion along the $x$-axis (orthogonal to the double trap) and only keep the parts when the particle is exclusively inside trap $A$ (resp. trap $C$). 
The PSD calculated from this sub time traces are fitted using equation~(\ref{eq:Sfit})  to determine the linewidth $\Gamma^\text{PSD}$ (See also section S2 where we did this for a particle trapped in a single well potential).
The obtained linewidths, for  the particle localised in both trap $A$ and trap $C$ are presented in Figure~\ref{fig:GamVsP}. We distinguish three interesting regimes.
\begin{itemize}
    \item For pressures below $10^3$~Pa, the intrinsic non-linearities of the trapping potential become dominant, and the PSD linewidth is not an appropriate measure of the damping $\Gamma$. Since we know that there is a linear relationship between the pressure and the damping $\Gamma$ and the pressure gauge is well calibrated in this regime, we can calculate the damping as $\Gamma = P^\text{gauge}/51$~Hz/Pa.
    \item For pressures between  $10^3$ and $5\times10^3$~Pa,the pressure gauge is still well calibrated and nonlinearities in the optical potential do not broaden the PSD. As a consequence, we use this regime to calibrate the proportionality factor between damping and pressure.
    \item Above $5\times10^3$~Pa, nonlinearities in the pressure gauge overestimate the actual pressure in the vacuum chamber. This leads to a deviation from the linear relationship. In contrast to the low pressure regime nonlinearities in the optical potential do not broaden the PSD and the fit to the PSDs provide a good estimate for the damping $\Gamma$.
\end{itemize}
Note that, since we are working with a double trap the non-linearities are higher than in the case of the single trap (see Figure~\ref{fig:calib}). Consequently, figure~\ref{fig:calib} and figure~\ref{fig:GamVsP} only agrees above  1000~Pa.\\

To summarize, we determine the experimental damping as 
$$
\Gamma = \left\{
    \begin{array}{ll}
        P_\text{gas}^\text{gauge}/51~\text{Hz/Pa}  & \mbox{for \ } P_\text{gas}<5\times10^3~\text{Pa} \\
        \Gamma_\text{PSD} & \mbox{for \ } P_\text{gas}\geq 5\times10^3~\text{Pa} \\
    \end{array}
\right.
$$
Conversely, the real pressure is given by 
\begin{equation}
    P_\text{gas}= 51~\text{Pa/Hz}~\Gamma\ .
\end{equation}
\quad \vspace{1em}

\subsection{Validity of the model}
Our quantitative analysis is limited by experimental drifts.
Drifts impact both, the shape of the potential and its measurement.

\begin{figure}[!h]
    \begin{center}
        \includegraphics[width=.8\textwidth]{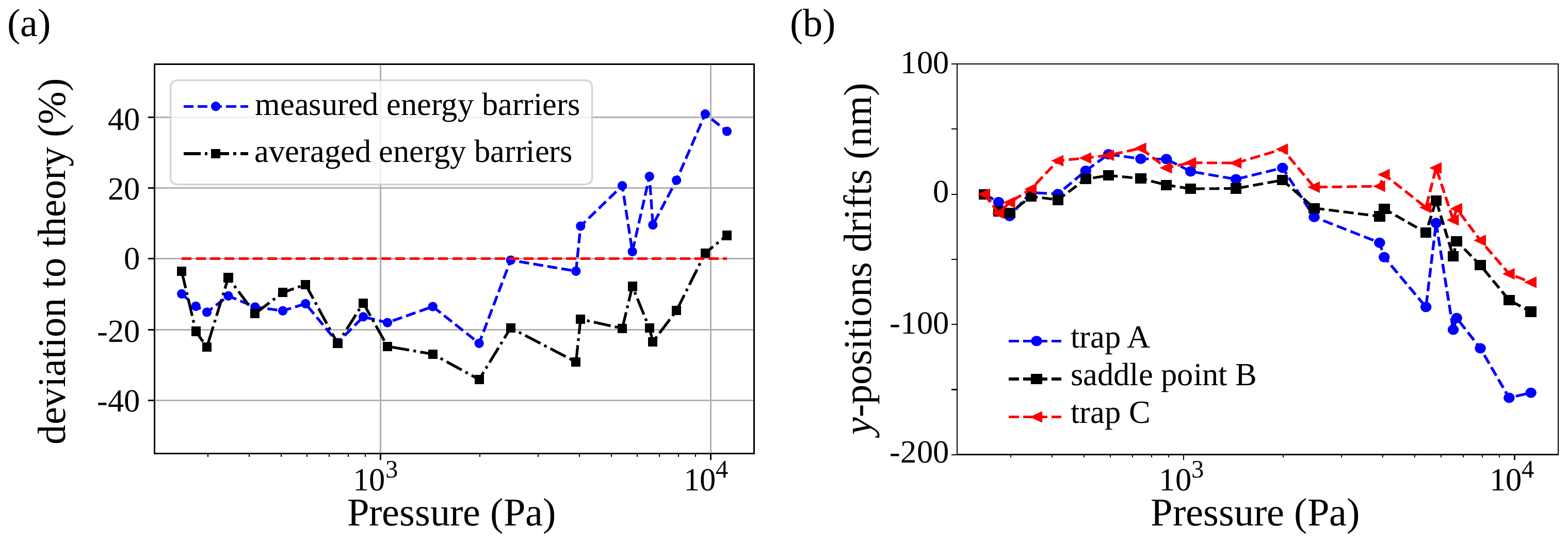}
    \end{center}
    \vspace{-1.5em}
    {\it    \caption{Errors and drifts. (a) Deviation between experimental rates, and the theoretical rate computed using equation~(3) of the main manuscript, with the potential parameters computed as the exact same pressure (blue), and with a fixed energy barrier (black). (b) Drifts of the trap and saddle point positions along the $y$-axis.  
    \label{fig:Errors}}}
\end{figure}

In the present case, we experimentally measure independently for each pressure the potential parameters, as previously discussed (see Methods and Figure~\ref{fig:PotCar}).
The estimation errors depends on:
\begin{itemize}
    \item the errors arising from fitting the potential. This error is estimated to be below 5~\% for the characteristic frequencies and a few percent for the energy barriers. 
    \item the inaccuracy of our non-linear measurement, which is much harder to quantify.   
\end{itemize}

To determine the impact of these two effects, we propose to compare, in Figure~\ref{fig:Errors}-(a) (blue curve), the error between the measured rate $R$ and the theoretically expected rate using the measured parameters at each pressure (Figure~\ref{fig:PotCar}).\\

At low pressure, in a regime where drifts  are negligible, we observe an almost constant error of less than $\approx 20$~\%. This error can be completely understood in terms of errors on potential parameters. In addition, the fact that this error is almost constant, is a strong indication for systematic errors in the estimation of potential parameters.\\ 

Above a few thousands Pascals, we observe that drifts start to play an increasing role. This is illustrated by the evolution of the trap and the saddle point positions, shown  in Figure~\ref{fig:Errors}-(b). These drifts, not only change the potential, but they also impact the measurement. For example, the regions of the potential that are further away from the center yields stronger non-linearities of the calibration functions (c.f.~Figure~\ref{fig:CalibXYZ}). These leads to an error in the estimation of the potential energy barriers. We believe that this effect is responsible for the fast change in the measured energy barriers at the highest pressures (c.f.~Figure~\ref{fig:PotCar}-(b)). To test this hypothesis, we compute the theoretical rate from the average energy barriers over the first ten pressures (up to 1050~Pa). This method reduces the difference between the theoretical and the experimental rate (Figure~\ref{fig:Errors}-(b), black). 
We therefore anticipate that improving the stability of the experiment should allow to measure the transition rates to within a few percent. 

Finally, note that the error in the measurement also affects the characteristic frequencies. However, this effect is weaker. Firstly because their measurement is local, in an area well calibrated, and secondly because errors in the energy barriers propagate more strongly because the rate depends exponentially on the energy barriers (see~Eq.~(1) of the main text).


\end{widetext}

\end{document}